\documentclass[10pt,a4paper]{scrartcl}

\usepackage{amsmath}  
\usepackage[usenames,dvipsnames]{color}
\usepackage{multirow} 
\usepackage{geometry}

\usepackage{amsfonts}
\usepackage{eurosym}
\usepackage{natbib}
\usepackage{a4}
\usepackage{epsf}
\usepackage{graphicx}
\usepackage{rotating}
\usepackage{tabularx}

\usepackage[latin1]{inputenc}
\newcommand{\beq}{\begin{equation}}
\newcommand{\eeq}{\end{equation}}
\numberwithin{equation}{section} 

\newcommand{\bm}[1]{\boldsymbol{#1}}

\usepackage{hyperref}
\usepackage{url}

\title{\textbf{STAD Research Report 2015/02} \\ \vspace{10mm} Parsimonious Time Series Clustering. \\   }

\author{Carmela Iorio*, Gianluca Frasso***, \\
Antonio D'Ambrosio*,Roberta Siciliano**\\
\\**Department of Economics and Statistics,
\\***Department of Industrial Engineering.
 \\University of Naples Federico II\\ 
\{carmela.iorio,antdambr,roberta\}@unina.it\\
\\****Institut des sciences humaines et sociales,\\
M\'ethodes quantitatives en sciences sociales\\
Universit\'e de Li\'ege, Belgium\\
gianluca.frasso@ulg.ac.be}

\begin{document}
\maketitle


\begin{abstract}
We introduce a parsimonious model-based framework for 
clustering time course data. In these applications the computational burden
becomes often an issue due to the number of available observations. 
The measured time series can also be very noisy and sparse and a suitable model 
describing them can be hard to define. We propose to model the observed 
measurements by using P-spline smoothers and to cluster the functional objects 
as summarized by the optimal spline coefficients. In principle, this idea can 
be adopted within all the most common clustering frameworks. In this 
work we discuss applications based on a k-means algorithm. We evaluate the 
accuracy and the efficiency of our proposal by simulations and by dealing with 
drosophila melanogaster gene expression data.

\end{abstract}
%
%
\section{Introduction}
\label{sec:introduction}
In many cases it is of interest to analyze phenomena
evolving over time. Time series data can be encountered, for examples, in 
biological, economical and engineering applications. 
Time series can be distinguished according to their nature (real or 
discrete valued, univariate or multivariate). Furthermore, these data can 
be collected over equal or different time points. In the second 
case the observed ``gaps'' are usually treated as missing values.

\begin{figure}
 \centering
 \includegraphics[width = 1\textwidth]{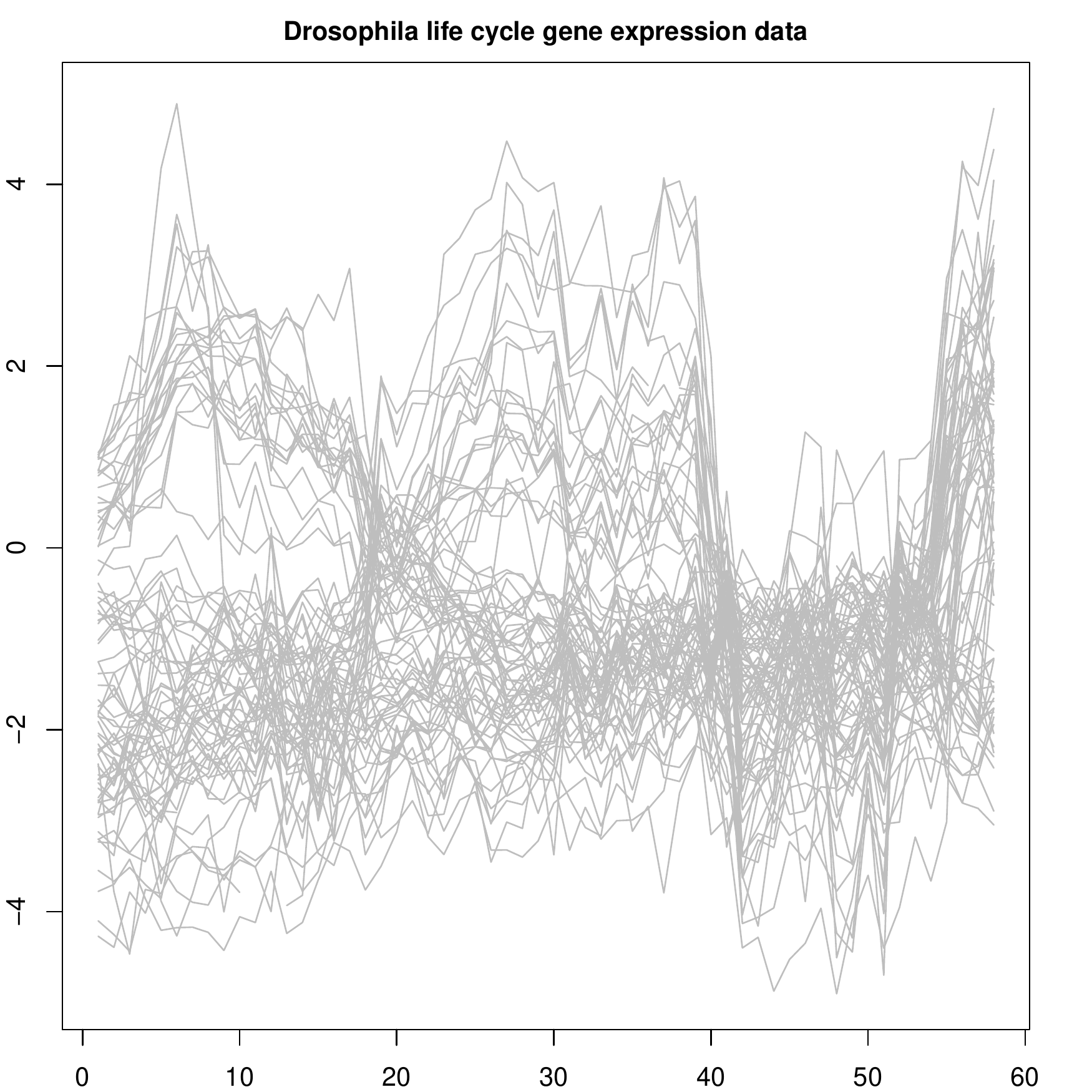}
 \caption{Subset of the drosophila melanogaster life cycle gene expression 
data.}
 \label{fig_raw_data}
\end{figure}

An example is presented in Figure~\ref{fig_raw_data} showing a subset of 
the drosophila melanogaster life cycle gene expression data analyzed in 
\cite{Arbeitman2002}. The original data set contains 77 gene expression profiles 
measured over 58 sequential time points from the embryonic, larval, and pupal periods 
of the life cycle. The gene expression levels were obtained by a cDNA microarray 
experiment.

Even if hardly identifiable by eyes, in Figure~\ref{fig_raw_data}, 
three classes of genes can be distinguished according to their 
gene expressions (see Section~\ref{sec:application}). Our 
interest here is to introduce a strategy to automatically partition these 
objects (time series) in homogeneous groups. This kind of
problem is often solved by defining a suitable ``clustering procedure''.
In the last years many techniques have been proposed to define homogeneous 
groups in time series data. A rich overview can be found in \citet{Liao2005}.

One of the first references on this topic is \citet{Maharaj2000} that proposed 
a two stage approach 
based on the hypothesis that the observed series could be modeled by invertible 
and stationary ARMA processes. A different point of view has been adopted by 
\citet{Baragona2001} which suggested a series of meta-heuristic approaches 
(e.g. simulated annealing, tabu search and the genetic algorithms) to partition 
stationary time series according the cross-correlations estimated from the 
residuals of models estimated on the original data. Self 
organizing map (SOM) algorithms have began popular clustering tools. An example 
of their application in longitudinal data analyzes is the method of 
\citet{Fu2001}.

\cite{Kumar2002} worked in a different direction and studied a new 
scale-invariant distance function based on the hypothesis of Gaussian 
distributed errors. In their model, time series sampled at $T$ points are 
represented by a sequence of probability distributions assumed independent
and identically normally distributed. \citet{Moller2003} introduced a 
short time series distance in order to measure the similarity between different 
series by measuring the relative change in the signal amplitudes.

Within a Bayesian framework, \citet{Ramoni2002} suggested to cluster univariate 
series by recursively merging the ones showing similar dynamics. In order to 
summarize a dynamic process by a transition probability matrix, each series is 
converted into a Markov Chain (MC) on which a clustering algorithm is then 
performed. This approach, known as Bayesian clustering by dynamics (BCD), adopts 
the posterior probability scoring metric to define the possible partitions 
searching for the one with the highest posterior probability through an 
entropy-based method. 

Many of the algorithms mentioned above do not facilitate the removal of the 
noise from data, encounter difficulties in handling data with missing 
values, require some pre-process of the series and do not account for possible
correlation in the measurement errors. Furthermore, when a large number of 
observations is considered, the computational effort related to the clustering 
task can become prohibitive. In order to overcome these hitches, many authors 
have suggested to exploit a flexible definition of the series functional form 
and to adopt some dimensionality reduction strategy. To these class belong the 
proposals based on the functional data analysis framework \citep{Ramsay2005}. 
The main idea is to describe the data by using optimal linear combinations of 
basis functions (e.g. by using B-spline bases) and to perform the clustering on 
the extracted signals or on the estimated basis coefficients (when possible). 

An example is the proposal of \citet{Chiou2007} exploiting a functional 
principal analysis of the raw series. \citet{Sangalli2010} 
suggested a functional k-means method to partition misaligned data series. 
\citet{James2003} introduced the {\it fclust} approach. Within this framework, 
the raw data are modeled by linear combinations of spline bases. The
regression spline coefficients are clustered taking them as 
distributed according to a mixture of Gaussian distributions with cluster 
specific means and common variance-covariance matrix. The spline bases are 
defined over equally spaced knots and their optimal number is chosen through 
cross-validation. A similar regression spline based approach has been proposed 
by \citet{Abraham2003} in combination with a k-means clustering procedure.

In a similar direction moves the work of 
\citet{Coffey2014}. The authors suggest to model the raw series by using 
truncated power P-spline smoothers \citet{Ruppert2003}. They exploit the 
connection between P-splines and linear mixed models to define a suitable 
variance-covariance cluster matrix to be used in a Gaussian mixture model. 
Within the Bayesian settings, a similar point of view has been adopted by 
\citet{Komarek2013} for clustering continuous and discrete longitudinal data.

In this paper we propose a parsimonious model-based approach for clustering 
longitudinal data. In particular we model the raw series by a linear 
combination of B-spline which coefficient are shrunk by difference penalties. 
This is the P-spline smoothing approach introduced by \citet{Eilers1996}. 
We model each series by P-splines and perform a cluster analysis on the 
optimal spline coefficients. This idea can be potentially adopted within
many of the most popular clustering methods but, in the present work, we 
focus our attention on the k-means one.

Our framework differs from the ones of \citet{James2003}, \citet{Abraham2003} 
and \citet{Coffey2014} due to the choice of the smoother. Indeed, B-spline 
based P-spline smoothers allow for a valuable simplification of the 
partitioning process. As discussed in \citet{Eilers2010} the P-spline 
coefficients represent the skeleton of the final fit. By summarizing the raw 
data using the estimated spline coefficients, one obtains an efficient reduction 
of the dimensionality of the partitioning task. This is not true if truncated 
bases P-splines are adopted (as in \citet{Coffey2014}). In addition, by 
shrinking the spline coefficients through difference penalties, the shape of 
the final smoothing function becomes quite insensitive to the choice of the 
number of bases and to their position allowing for efficient data interpolation 
in the case missing values (this is an advantage with respect to the proposals 
of \citet{James2003} and \citet{Abraham2003}).

This paper is organized as follows. Section~\ref{sec:p_splines} reviews the 
P-splines smoothing framework discussing briefly the main features that will 
be useful for our applications. In section~\ref{sec:cluster} we introduce our 
proposal within the k-means clustering framework. In 
Section~\ref{sec:simulations} the performances of our method are evaluated 
through simulation while in Section~\ref{sec:application} the gene expression 
data of Figure~\ref{fig_raw_data} are analyzed. Section~\ref{sec:discussion} 
concludes the paper with a discussion on our achievements and possible 
future research guidelines.
%
%
\section{P-splines in a nutshell} 
\label{sec:p_splines}
P-splines have been introduced by \citet{Eilers1996} as flexible smoothing 
procedures combining B-spline bases \citep[see e.g.][]{deBoor1978} and 
difference penalties.

Suppose to observe a set of data $\{x,y\}_{j=1}^{n_i}$ 
where the vector $\bm{x}$ indicates the independent variable (e.g. time) 
and ${\bm y}$ the dependent one. Our aim is to describe the available 
measurements through an appropriate smooth function. We assume that the 
observed series can be modeled as:
\begin{equation}\label{eq:general_model}
 \bm y = f(\bm x) + {\bm \epsilon}, 
\end{equation}
where $\bm \epsilon$ is a vector of errors and $f(\cdot)$ is an unknown 
smooth function.

Denote with $\bm{B}_{h}(x;q)$ the value of the $h$th B-spline of degree $q$ 
defined over a domain spanned by $m$ equidistant knots (in what follows we 
consider always equally spaced knots as suggested by \citet{Eilers1996, 
Eilers2010}). A curve that fits the data is given by $\hat{y}(x) = \sum_{h} 
a_{h} \bm{B}_{h}(x;q)$ where $a_{h}$ (with $h = 1,...,m + q$) are the 
B-splines coefficients estimated through least squares. Unfortunately the 
curve obtained by minimizing $\|{\bm y} - {\bm B \bm a}\|^{2}$ w.r.t. $\bm{a}$ 
shows more variation than is justified by the data if a dense set of spline 
functions is used. In order to avoid overfitting one can estimate ${\bm{a}}$ in 
a penalized regression setting:
\begin{equation}\label{eq:pen_reg}
\hat{{\bm a}} = \mathop{\mbox{argmin}}_{{\bm a}} \|{\bm y} -  
{\bm B \bm a} \|^{2} + \lambda\|{\bm D_{d} \bm a} \|^{2},
\end{equation}
where $\bm{D}_{d}$ is a $d$th order difference penalty matrix such that 
${\bm D_{d} \bm a} = \Delta^{d}{\bm a}$ defines a vector of $d$th 
order differences of the coefficients. Usually $d = 2$ or $d = 3$ is used. A 
second order difference matrix appears as follows:
\begin{equation*}
{\bm D}_{2} =
\left[
 \begin{array}{cccccc}
1&-2&1& \cdots & 0&0\\
0&1&-2&1&\cdots &0\\
\vdots& \ddots & \ddots & \ddots &\ddots & \vdots \\
0&0& \cdots &1&-2&1\\
  \end{array}
\right].
\end{equation*}
The optimal spline coefficients follow from \eqref{eq:pen_reg} as:
\begin{equation}\label{eq:spline_coefficients}
\hat{{\bm a}} = ({\bm B}^{\top} {\bm B} + \lambda {\bm D}^{\top} 
{\bm D})^{-1} {\bm B}^{\top} {\bm y}.
\end{equation}

The (positive) parameter $\lambda$ controls the degree of smoothness of the 
final fit. For a smoothing parameter close to zero the final smoother tends to 
interpolate the observations (leading to overfitting) while, for large values 
of $\lambda$, the fitted function will tend to a polynomial of degree $d-1$.

\begin{figure}
\centering
\includegraphics[width=1.00\textwidth]{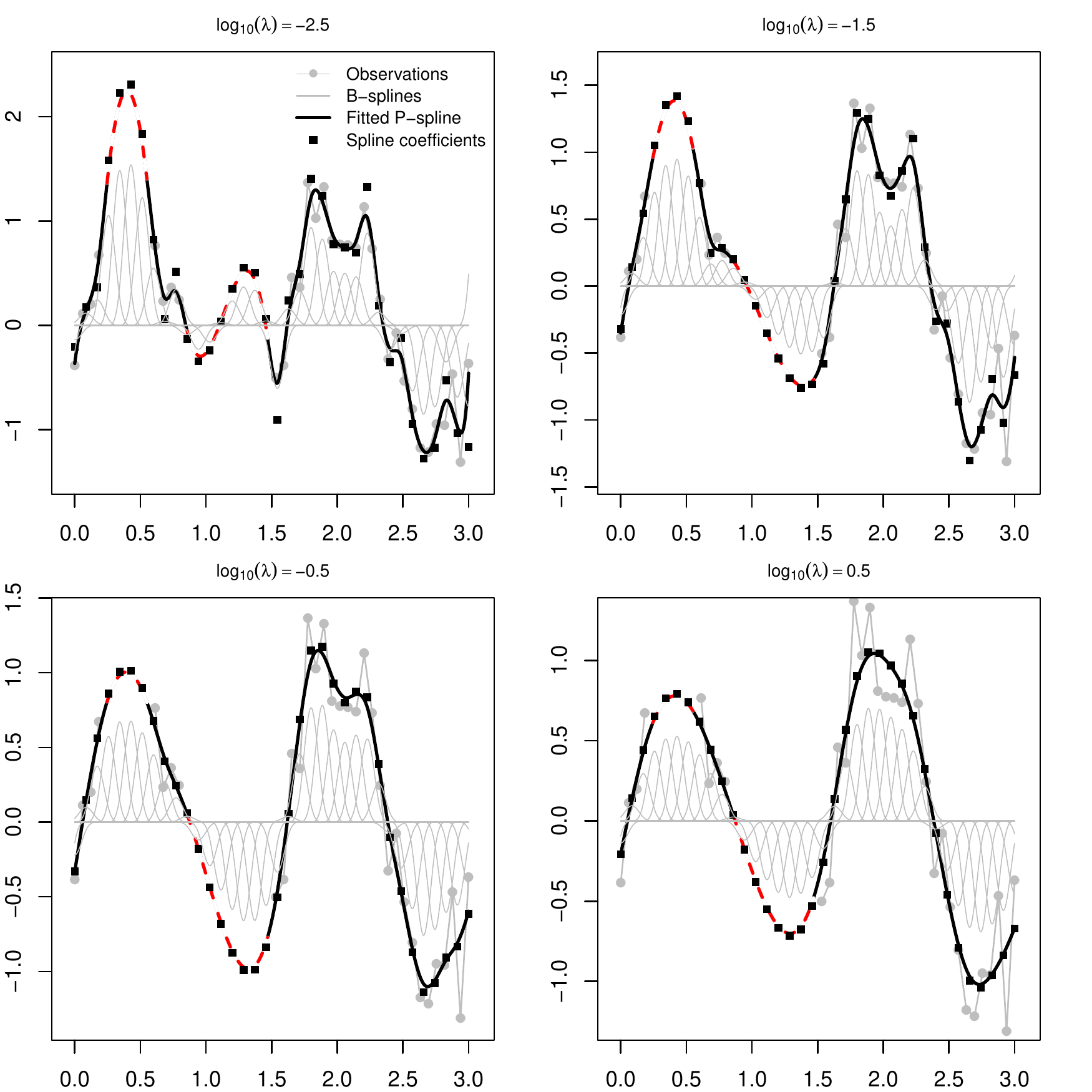}
\caption{Smoothing and interpolation with P-splines (solid black lines) for 
different values of the $\lambda$ parameters (in a $\log_{10}$ scale). The dashed red 
segments represent the estimated signals in correspondence of missing data. 
The squared dots indicate the estimated optimal spline coefficients. The 
scaled basis functions are shown at the bottom of each graph.}
\label{fig_simul_p_splines_lambda}
\end{figure}

Figure~\ref{fig_simul_p_splines_lambda} shows the estimated P-splines for 
different values of the smoothing parameters (for brevity only four $\lambda$ 
values are considered). The data have been simulated by adding a Gaussian noise 
to a sinusoidal signal (50 observations). Large portions of observations have 
been omitted to simulate the presence of missing values. Cubic B-splines defined 
on 30 equidistant knots and third order penalties have been used.
As it appears form Figure~\ref{fig_simul_p_splines_lambda}, the spline 
coefficients (squared dots) represent the skeleton of the fitted smoother. It 
can be shown that the coefficients of interpolating P-splines define a 
polynomial sequence of degree $2d-1$ and that, thanks to the penalty, they are 
forced to follow a smooth pattern also when some observations are missed. 

\begin{figure}
 \centering
 \includegraphics[width = 1\textwidth]{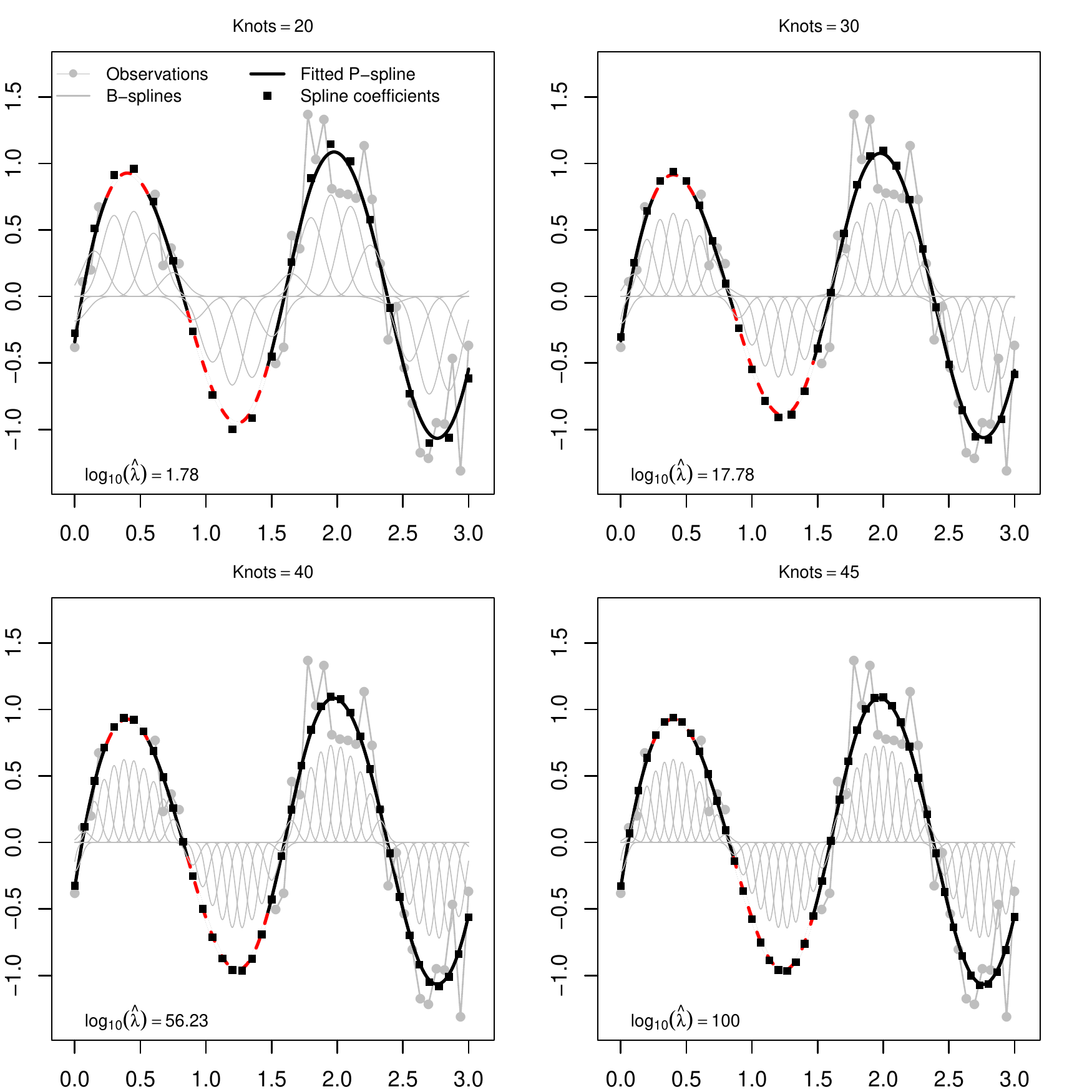}
 \caption{Smoothing and interpolation with P-splines (solid black lines) for 
different numbers of B-spline bases defined on equally spaced knots. The value 
of the optimal smoothing parameters have been selected using the V-curve 
criterion. The dashed red segments represents the estimated signals in correspondence 
of missing data. The squared dots indicates the estimated optimal spline 
coefficients. The scaled basis functions are shown at the bottom of each 
graph.}
 \label{fig_simul_p_splines_bases}
\end{figure}

The selection of the optimal number of basis functions and their 
location are crucial in spline regression 
applications and greatly influence the final fit. P-splines make these choices 
not particularly relevant. Indeed, the shape of the final smoother is mostly 
governed by the smoothing parameter and is hardly influenced by the richness of 
the basis matrix and by the knot location. This is due to the characteristics 
of the penalty and of the basis functions. 
Figure~\ref{fig_simul_p_splines_bases} investigates this aspect where the optimal 
smoothing parameters have been selected through the V-curve procedure 
introduced in Section~\ref{subsec:smoothing_selection}. For this reason, we 
perform all our analyzes using smoothers built over a generous number of 
equally spaced knots (but still smaller than the available observations). 

For a complete discussion about the properties of the P-spline smoother 
we refer to \citet{Eilers1996, Eilers2010}. These features are not shared
by other definitions of P-splines 
(e.g. the truncated power P-splines exploited by \citet{Coffey2014}) and
by regression splines (suggested by \citet{Abraham2003} and \citet{James2003} for
functional clustering applications) and play a crucial role in the method 
described in this paper. 
%
%
\subsection{Smoothing parameter selection}
\label{subsec:smoothing_selection}
From Figure~\ref{fig_simul_p_splines_lambda} it clearly appears that the shape 
of the fitted functions depends on the value of the smoothing parameter. For 
this reason it is of great interest to select the optimal $\lambda$ by a
reliable automatic procedure. The Akaike information criterion, the 
Bayesian information criterion and the (generalized) cross validation 
are popular alternatives.

These ``classical'' methods suffer of two drawbacks: 1) they require the 
computation of the effective model dimension (see e.g. \citet{Hastie1990}) which 
can become time consuming for a rich set of B-spline functions, and 2) they 
are sensitive to serial correlation in the noise around the trend (leading to
overfitting). This last aspect is crucial in many applications. 
For this reason, in all the examples presented in this paper, we have adopted 
the V-curve criterion to select the optimal smoothing parameter. 

The V-curve can be viewed as a convenient simplification of 
the L-curve framework proposed by \citet{Hansen1992}.
The L-curve is a parameterized curve comparing the two ingredients of every 
regularization or smoothing procedure: badness of the fit and roughness of the 
final estimates. Given a P-spline defined for a fixed $\lambda$, the 
following quantities can be computed:
  \begin{equation*}
  \{\omega(\lambda); \theta(\lambda) \} = \{ \|{\bm y} - {\bm{B}} 
\hat{{\bm {a}}}(\lambda)\|^{2}; \|{\bm {D} }  \hat{{\bm 
{a}}}(\lambda) \|^{2}\}.
  \end{equation*}
The L-curve is the plot of $\psi(\lambda) = \log(\omega)$ against 
$\phi(\lambda) = \log(\theta)$ evaluated over a grid of smoothing parameters. 
This plot typically shows a L-shaped curve and the optimal amount of smoothing 
is located in the corner of the ``L'' by maximizing a measure of the local 
curvature.

The V-curve criterion simplifies this selection criterion by requiring the 
minimization of the Euclidean distance between the adjacent points lying on the 
L-curve. The optimal smoothing parameter is then the one minimizing $V(\lambda) 
= \sqrt{ \{ \Delta \psi(\lambda) \}^{2} +  \{ \Delta \phi(\lambda) \}^{2}}$ 
($\Delta$ indicates the first order difference operator).
In this way the computation of the partial second derivatives needed 
for the evaluation of the local curvature is avoided (see \citet[][]{Frasso2015} 
for a detailed discussion). 
%
%
\section{P-spline based k-means procedure}
\label{sec:cluster}
Any cluster analysis requires two choice: the identification of a 
suitable clustering algorithm and the selection of an appropriate distance 
measure. In what follows, we introduce the notations, the clustering algorithm 
and the distance measures used in the paper.

\subsection{Notations and definitions}
Let $\bm y_i$ be the ${i}$th observed series with $i = 1, \ldots, N$. 
Let $\bm c_{k}$ be the $k$th cluster center, with  $k=1, \ldots, K$ and $1 < K < N$.
Define as $\bm a_{i}$ the $[(m + q) \times 1]$ vector of optimal spline 
coefficients estimated for the $i$th series. Finally, in what follows, 
we indicate with $x_j$, $j=1,\ldots, n_i$, the time points 
over which the $i$th series is observed.

\subsection{Clustering algorithm and distance measures}
The k-means algorithm (see e.g. \citet{Mcqueen1967} and 
\citet{Hartigan1979}) is one of the most popular clustering approaches. It 
relies on an iterative scheme. The procedure aims to partition 
the observations in a predetermined number ($K$) of clusters. In the case the 
algorithm is applied to the P-spline coefficients estimated for $N$ time series, 
its steps can be summarized as follows:
\begin{itemize}
\item[(1)] Initialization: fit a P-spline smoother to each series and store the 
optimal spline coefficient in a matrix $\bm{A} = [\widehat{a}_{i,h}]$ of 
dimension $[(m + q) \times N]$. Assign randomly the columns of $\bm A$ to $K$ 
groups representing the initial clusters. 
 
\item[(2)] Assign each sequence $\bm a_{i}$ to the cluster whose distance from 
the center is minimum.

\item[(3)] When all objects have been assigned to a group, update the positions 
of the $K$ cluster centroids.
\end{itemize}
The procedure is stopped when the cluster centers do not move any more. 
Otherwise steps 2 and 3 are repeated until convergence.

The $K$ centroids $({\bm c_{k}|k=1,...,K})$ are defined as the minimizers 
of:
\begin{eqnarray}\label{eq:distance_general}
J = \sum_{i=1}^{N} \sum_{k=1}^{K} \mathcal{D}(\bm a_{i}, \bm c_{k})
\end{eqnarray} 
where $\mathcal{D}(\bm a_{i}, \bm c_{k})$ is a distance (dissimilarity) measure 
between the $i$th spline coefficient vector and the $k$th cluster center.
In what follows we focus on two distances: the Euclidean distance and the 
distance based on the Pearson's correlation coefficient.

The Euclidean distance is computed as follows:
\begin{equation}\label{eq:euclidean_distance}
 \mathcal{D}_{2} = \sum_{i=1}^{N} \sum_{k=1}^{K} \| \bm a_{i} - \bm c_{k} 
\|^2
\end{equation}

In time series applications, a distance measure based on the Pearson's 
correlation coefficient is often a valid alternative to 
\eqref{eq:euclidean_distance}:
\begin{equation}
\mathcal{D}_{\rho} = 1 - \rho(\bm a_{i}, \bm c_{k}),
\end{equation}
where $\rho(\cdot)$ is the Pearson's correlation coefficient.

The mentioned metrics are defined in the time domain. Distances defined in the 
frequency domain can be more appropriate to measure the similarity 
between time series in particular applications. We refer to \citet{Vilar2010} 
and \citet{Montero2014} for a  detailed discussion about possible distance 
measures applicable in time series clustering tasks.

Finally notice that, in the case of normally distributed errors in 
Eq.~\ref{eq:general_model}, the asymptotic properties of the k-means algorithm 
described above can be proven following the lines discussed in 
\citet{Abraham2003}.  

\section{Simulations}
\label{sec:simulations}
In order to test the performances of our proposal we present here the results 
of a simulation study. 

We have generated $K=6$ clusters of series observed over $100$ time points 
with $x_{j} \in [0,1]$. Each cluster is formed by adding an error term to a specific
signal functional form. We defined the following functional classes:
\begin{eqnarray}
\nonumber &\mbox{Sin}\ & y_{ij}^{(1)} = \alpha_{i} \sin(4 \pi x_{ij}) + 
\beta_{i} + \epsilon_{ij} \\
\nonumber &\mbox{Cubic}\ & y_{ij}^{(2)} = \alpha_{i} (x_{ij} + 0.73)^{3} + 
\beta_{i} + \epsilon_{ij}\\
\nonumber &\mbox{Neg-pow}\ & y_{ij}^{(3)} = \alpha_{i} (x_{ij} + 0.5)^{-3/2} + 
\beta_{i} + \epsilon_{ij}\\
\nonumber &\mbox{Cos}\ & y_{ij}^{(4)} = \alpha_{i} \cos(2 \pi x_{ij})
 + \beta_{i} +\epsilon_{ij}\\
\nonumber &\mbox{Exp}\ & y_{ij}^{(5)} = \alpha_{i} \exp(- 6 x_{ij}) + 
\beta_{i} + \epsilon_{ij}\\
\nonumber &\mbox{Lin}\ & y_{ij}^{(6)}= - \alpha_{i} (x_{ij} - 0.5) + 
\beta_{i} +\epsilon_{ij},
\nonumber
\end{eqnarray}
where $\alpha_{i} \sim {\cal N}(4, \sigma^{2}_{\alpha})$ and $\beta_{i} \sim 
{\cal N}(0, \sigma^{2}_{\beta})$ with $\sigma_{\alpha}$ and $\sigma_{\beta}$ 
drawn from ${\cal U}(0.3, 1)$.

We have taken into account three possible scenarios for the error component 
$\bm \epsilon_{i}$: \textit{scenario} 1) a 
zero mean normally distributed noise with standard deviation $\sigma_{\epsilon} 
\sim {\cal U}(0, 0.5)$; \textit{scenario} 2) a first order autoregressive model with 
$\sigma_{\epsilon} \sim {\cal U}(0, 0.5)$ and correlation coefficient $\rho = 
0.5$; \textit{scenario} 3) a similar autoregressive process with $\rho = 0.9$. 

For each class have been generated respectively 90, 50, 100, 
25, 60 and 35 series. Each series have been modeled by P-splines taking 
cubic bases and third-order penalties. The optimal smoothing parameters 
have been selected through a V-curve procedure (introduced in 
Section~\ref{sec:p_splines}). 

\begin{figure}
\includegraphics[width=0.9\textwidth, angle = 90]{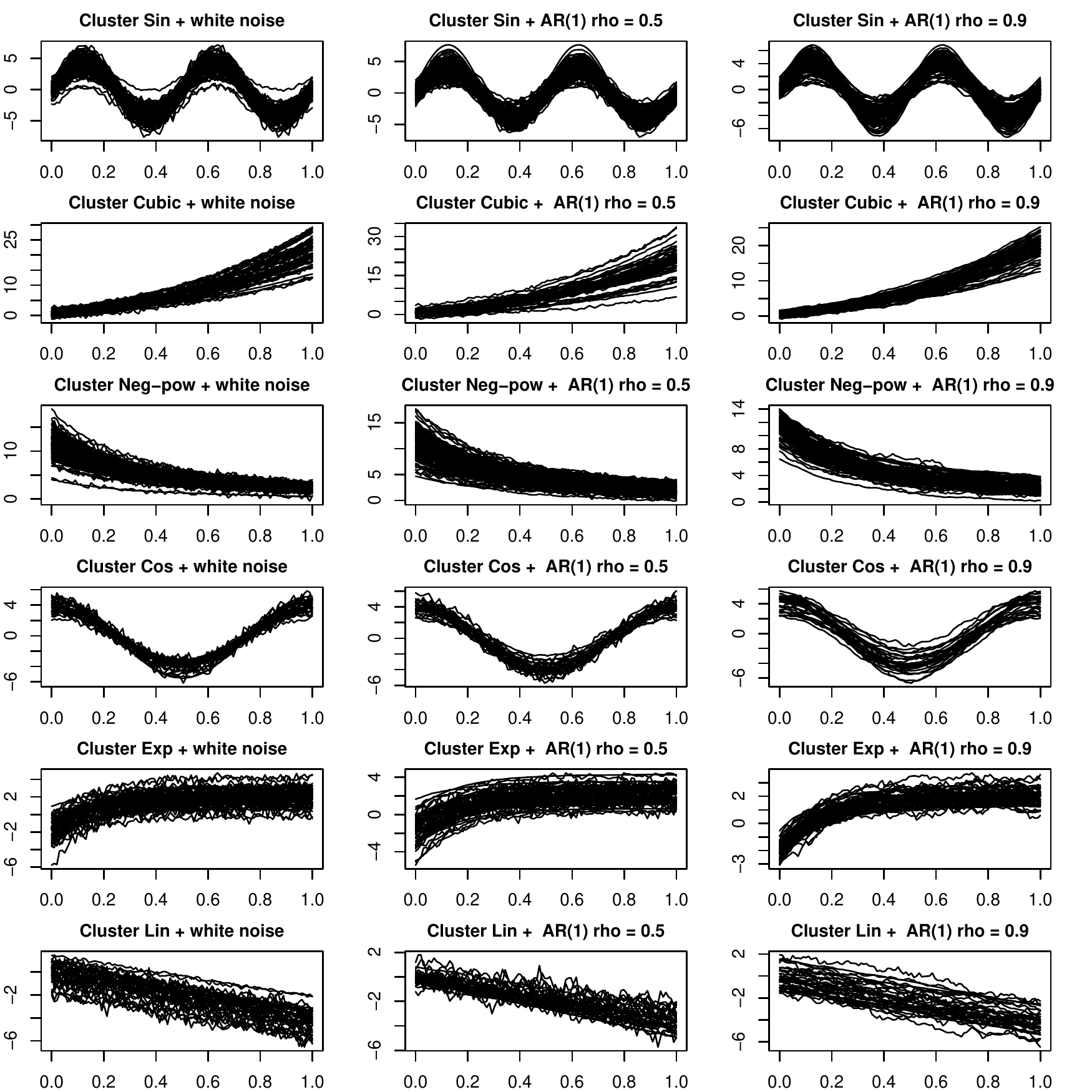}
\caption{Example of simulated data without missing observations.}
\label{simulation_settings_complete}
\includegraphics[width=0.9\textwidth, angle = 90]{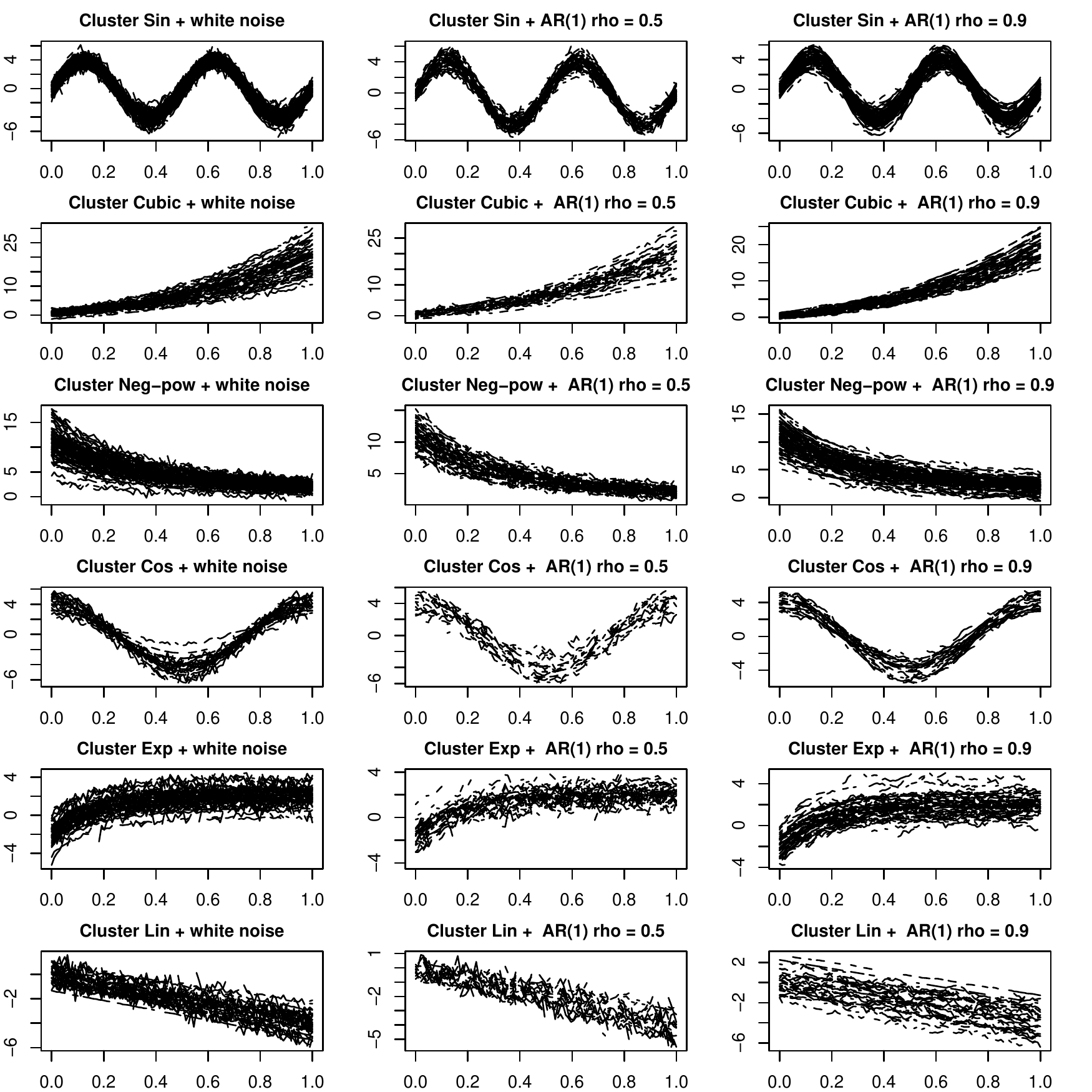}
\caption{Example of simulated data with missing observations.}
\label{simulation_settings_missing}
\end{figure}

Moreover, we have considered both complete and incomplete 
observations. In the latter case, the missing mechanism has been simulated by 
excluding independently and uniformly some observations from each series. The 
percentage of missing values for each series has been supposed uniformly 
distributed between 10\% and 50\%.
Figure~\ref{simulation_settings_complete} and Figure~\ref{simulation_settings_missing} show
some simulated datasets.

For this simulation study we used a k-means algorithm based
on the Pearson's correlation distance. The procedure was run taking 50 random starting
point with 50 replicates. To check the sensitivity of our procedure with respect to a different
number of cubic splines, we performed the analyzes by taking
the number of internal knots equal to the 10\%, 20\%, 30\%, 40\% of the 
observations.
In order to evaluate the clustering performances 
we computed the Adjusted Rand Index (ARI) validation criterion 
\citep{Hubert1985}. The results have been obtained over 100
replicates per each scenario. 

We compare the performances of the proposed approach with those achieved
by two frameworks based on the k-means partitioning: one exploiting 
regression splines \citep[see e.g.][]{Abraham2003} and 
the functional PCA k-means approach of \citet{Chiou2007}.
%
%
\subsection{Simulation results with complete observations}
\label{subsec:simul_complete}
\begin{figure}
\centering
\includegraphics[width=1.00\textwidth]{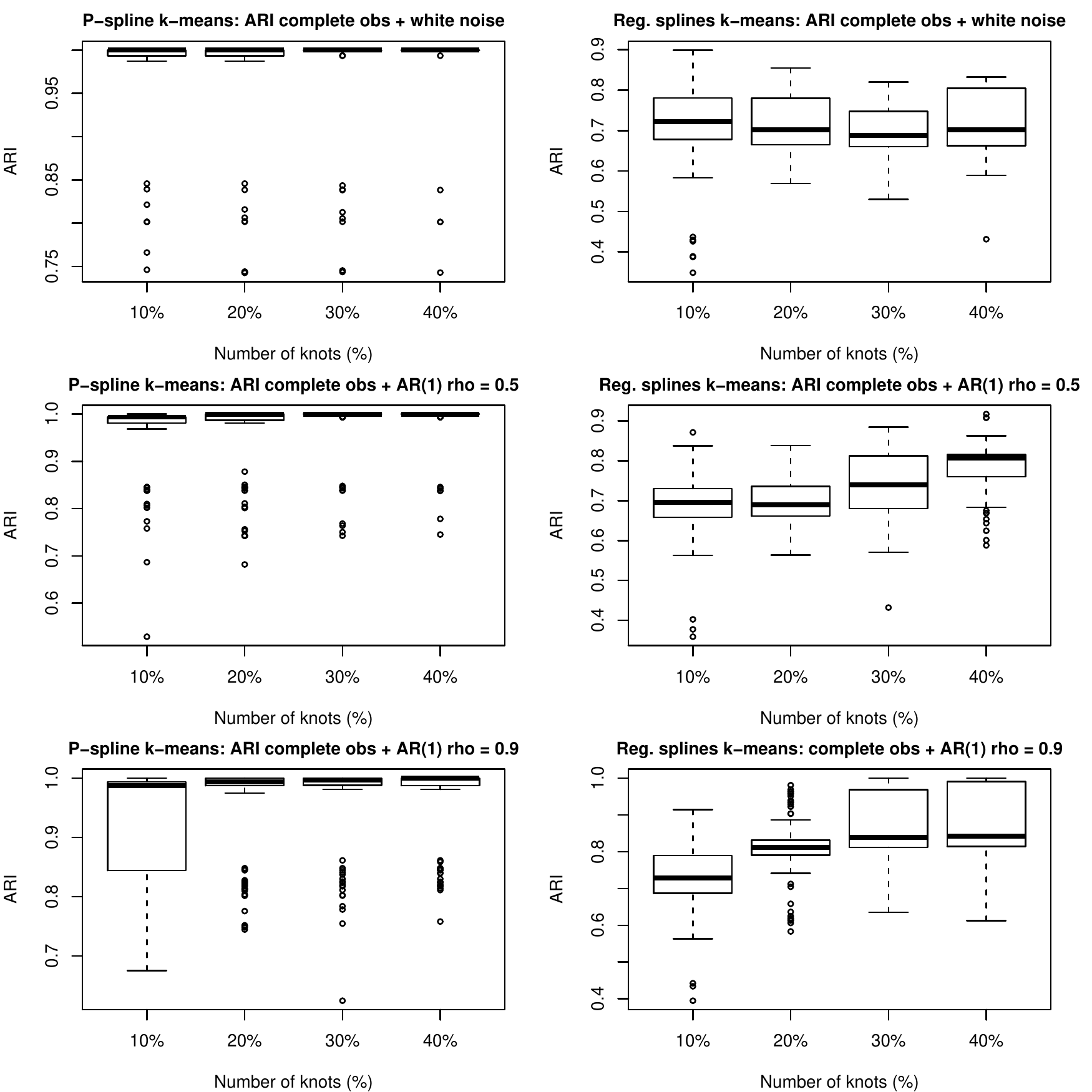}
\caption{Clustering results for simulated series with complete observations. 
The panels on the left column present the box-plots of the ARI values
registered for the P-spline based approach for different numbers of spline functions and
different definitions of the noise term. The results obtained by using a k-means partitioning 
of the regression spline coefficients are shown in the right column.}
\label{fig_sim_complete_cases}
\end{figure}

Figure~\ref{fig_sim_complete_cases} shows the box-plots of the ARI
values for the P-splines based k-means approach 
(left column) and for a k-means clustering based on
regression splines coefficients (right column). In both cases
we have used cubic B-splines defined over equally spaced knots.
The performances have been evaluated considering complete observations. 
In each cell of the figure matrix four box-plots are shown 
according to the different number of spline functions involved 
in the smoothing procedure.

The clustering performances of the P-spline based k-means method seem 
quite insensitive to the choice of the number of internal knots. 
Even with cubic B-splines defined over 10 equally spaced interior knots, 
the mean ARI was found equal to $0.98$, $0.96$ and $0.94$ for scenario 1, 2 and 3, 
respectively. As expected, the variability of the ARI distributions
seems to be influenced by the number of knots used to build the P-splines smoothers
(especially for relatively small number of spline functions).
Finally, regardless the correlation of the error component,
the mean ARI values grow (slightly) with the number of internal knots 
even if it was found close to $1$ for all the settings. 

The regression spline based procedure shows
a lower classification quality in all the simulation settings. This can be explained 
by the fact that, without the introduction of a smoothing penalty,
the estimated spline coefficients (and then the final fits)
are strongly influenced by the number and the position of the knots used
to define the basis functions. These aspects are not crucial for
P-spline smoothers as discussed in Section~\ref{sec:p_splines}.

As expected, the computational cost of the two procedures tends to be larger
as the number of bases increases. Nevertheless, for the P-spline based 
classification, the average times  required by using cubic B-splines defined
over 10 equally spaced interior knots was found equal to $2.5$, $2.8$ and $3$ seconds 
for scenario 1, 2 and 3, respectively and increases (approximately) proportionally 
with the number of spline knots. The computational burden for 
the k-means partitioning based on regression spline coefficients is significantly lower 
given that the selection of the smoothing parameter is not required.

These results can be compared with the one achieved by the functional
PCA k-means approach proposed by \citet{Chiou2007} (upper panel of
Figure~\ref{fig_sim_results_functional_PCA}). This approach ensures
performances close to the ones registered by the P-spline based k-means
and appears robust to the characteristics of the error terms.

%
%
\subsection{Simulation results with missing observations}
\label{subsec:simul_missing}
\begin{figure}
\centering
\includegraphics[width=1.00\textwidth]{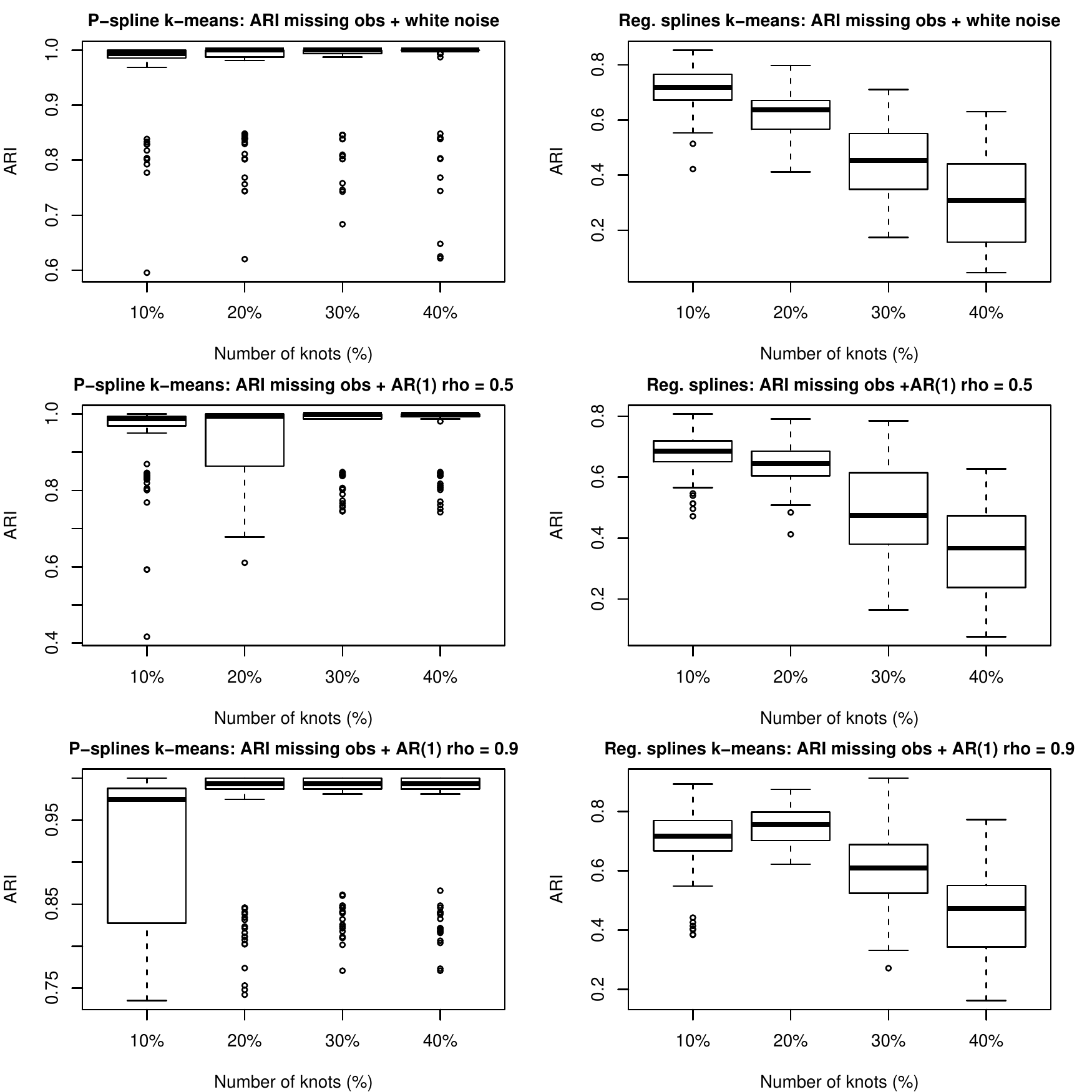}
\caption{Clustering results for simulated series with missing observations. 
The panels on the left column present the box-plots of the ARI values
registered for the P-spline based approach for different numbers of spline functions and
different definitions of the noise term. The results obtained by using a k-means partitioning 
of the regression spline coefficients are shown in the right column.}
\label{fig_sim_missing_cases}
\end{figure}

Figure~\ref{fig_sim_missing_cases} shows the performances of 
the P-splines based k-means approach 
(left column) and of the k-means clustering procedure based on
regression splines coefficients (right column) in presence of missing
observations. The box-plots in each panel summarize the distribution of 
the ARI values computed for different number of spline functions.

For the P-spline based k-means procedure, the average ARI values were found lower than
the ones obtained for the complete observation case but still above above 95\% for
all the simulation settings. On the other hand, the presence of missing observations
have a strong impact on the performances of the regression spline k-means method. 
This difference between the two approaches can be easily explained by considering
the interpolation properties guaranteed by the finite difference penalty matrix involved
in the definition of the P-spline smoothers.

As expected also in this case an increasing number of knots requires a larger computational effort,
whatever the scenario for the error component. For the P-spline based k-means procedure, 
the average time required by using 40 equally spaced interior knots was equal 
to $7.07$, $7.08$ and $7.5$ seconds for scenario 1, 2 and 3, respectively.

\begin{figure}
\centering
\includegraphics[width=1.00\textwidth]{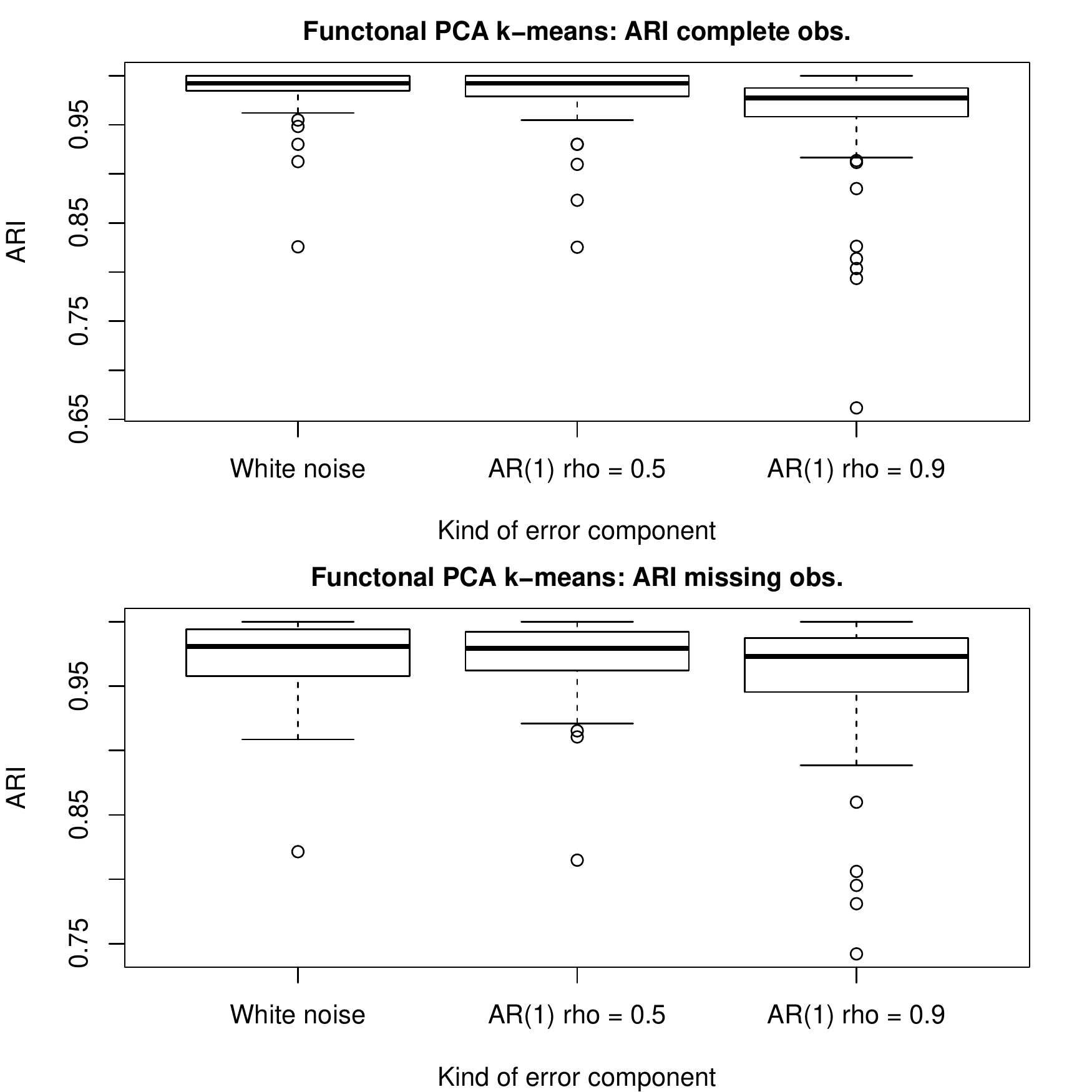}
\caption{Clustering performances of the functional PCA k-means approach. 
The upper panel shows the distribution of the ARI
values registered for complete datasets while the lover plot 
indicates the ones observed for datasets with missing
observations.}
\label{fig_sim_results_functional_PCA}
\end{figure}

Finally, these results can be compared with those achieved by the functional PCA k-means approach
(lower panel of Figure~\ref{fig_sim_results_functional_PCA}). As before, this method ensures 
performances closer to the ones registered for the P-spline based k-means framework. 


%
%
\section{Clustering drosophila gene expression data}
\label{sec:application}
\begin{figure}
\centering
\includegraphics[width=1.00\textwidth]{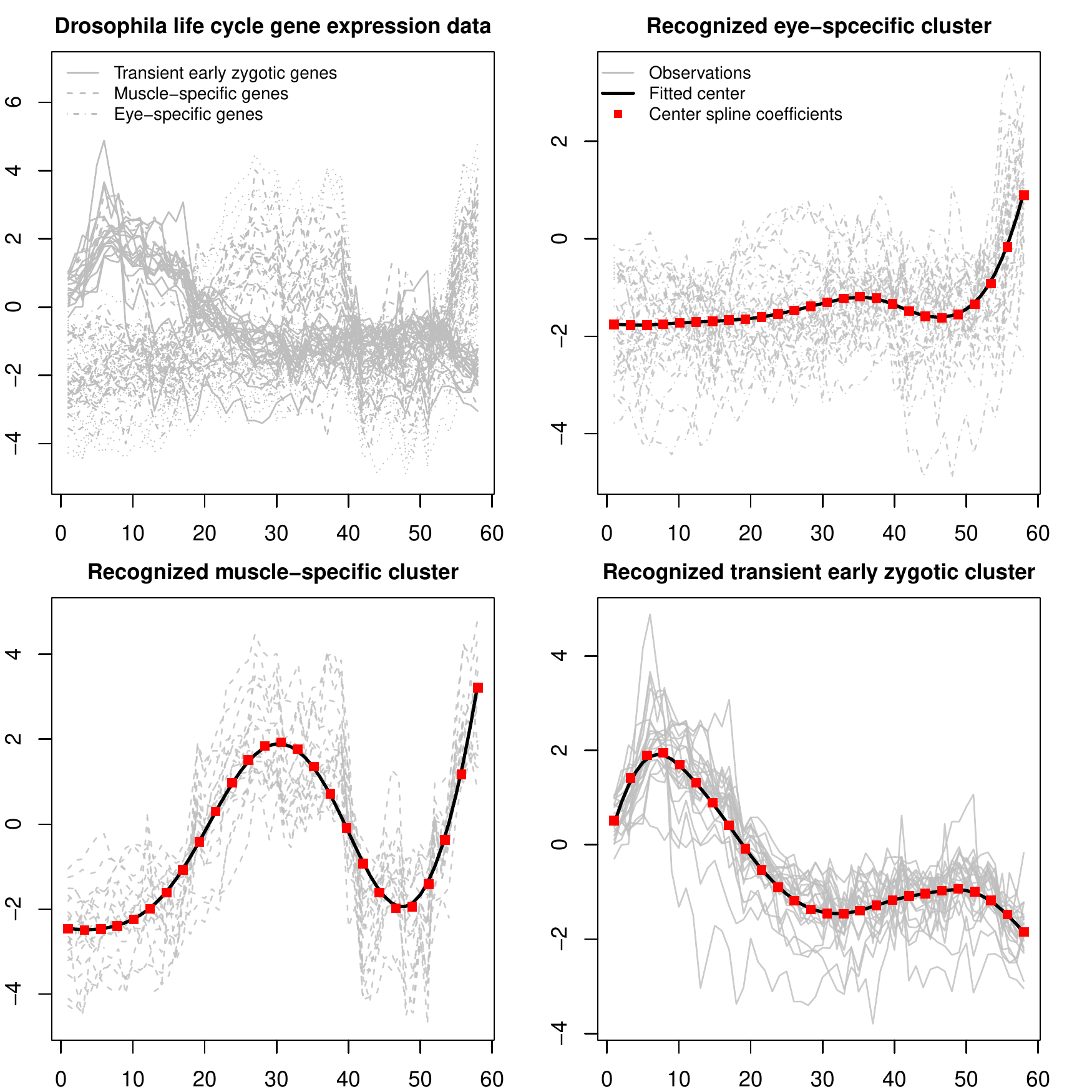}
\caption{Gene expression profiles clustered by k-means procedure with Euclidean 
distance. For each subplot the horizontal axis represents the time and the 
vertical axis the gene expression level. The gray lines reproduce the raw data 
series assigned to each cluster. The red dots indicate the estimated optimal 
spline coefficients for cluster center. The black solid lines indicate the 
center functionals.}
\label{fig_gene_expression}
\end{figure}

In this section we analyze the data introduced in 
Section~\ref{sec:introduction}. As already mentioned,
\citet{Arbeitman2002} have identified three gene categories according to their 
biological functions. In the data of Figure~\ref{fig_raw_data}, it is possible 
to identify $23$ muscle-specific, $33$ eye-specific and $21$ transient early 
zygotic genes (upper left panel of Figure~\ref{fig_gene_expression}). Our aim 
here is to test the capability of the P-spline based k-means procedure introduced 
above in identifying these groups.

In order to partition the data we used a k-means algorithm based on the 
Euclidean distance. We set $K=3$ possible clusters. The analysis has been 
performed by taking $100$ replicates of the partitioning procedures in order to 
avoid local minima. Each gene expression profile has been smoothed by P-splines. 
We used third order penalties and cubic B-splines defined over 25 equally 
spaced interior knots.

Figure \ref{fig_gene_expression} shows the results of our clustering proposal. 
The recognized center are really similar to the ``true'' ones defined in 
\citet{Arbeitman2002}. 

The entire analysis has been performed in less than a second. 
The adjusted rand index, computed on the basis of the groups identified by 
\citet{Arbeitman2002}, was found approximately equal to $0.96$. The goodness of 
our results can also be appreciated by looking at the homogeneity of the 
functions assigned to each cluster. Furthermore, these results are consistent 
with those presented by \citet{Chiou2007}.   

In this application, as in the simulation exercise presented above, we did not 
select the optimal number of clusters taking it as known. However, all the 
cluster selection procedures proposed in the literature for k-means 
partitioning are in principle applicable to the discussed method.
%
\section{Discussion}
\label{sec:discussion}
In this paper we have presented a parsimonious clustering method 
suitable for time series applications. The idea behind our proposal is quite 
simple but efficient. We suggest to model each series by P-spline smoothers as 
defined by \citet{Eilers1996} and to perform a cluster analysis on the 
estimated coefficients. This makes possible to summarize the observed 
series in a lower-dimensional vector of parameters. 

As briefly discusses in Section~\ref{sec:p_splines}, P-splines show a series of 
desirable properties that make the introduced procedure 
particularly attractive (a detailed discussion is presented by 
\citet{Eilers2010}). First of all, the optimal 
P-spline coefficients are close to the fitted curve and represent the skeleton 
of the final fit. Second, the final estimates are hardly influenced 
by the number and location of the knots used to define the B-spline bases 
and depend mainly on the weight assigned to the roughness penalty. Finally, the 
combination of B-spline functions and difference penalty ensure an efficient 
interpolation of the observed measurements also when some of them are missed. 

In order to select the optimal amount of smoothing to apply to each series we 
suggested to use a V-curve approach. Even if alternative 
criteria (such as AIC, BIC and cross validation) can be adopted, the V-curve 
appears particularly appealing from a computational point of view and ensures 
robustness against possible correlation in the noise of the observed data (we 
refer to \citet{Frasso2015} for a complete discussion about these aspects).

In this work we have chosen 
to classify the estimated spline coefficients by k-means algorithm.
The performance of our P-spline based k-means approach have been evaluated by 
intensive simulations. Our simulation study has shown that, both in presence of 
complete or missing observations, our proposal ensures high quality 
performances in a reasonable computational time regardless the complexity (in 
terms of number of spline functions) of the adopted P-spline models. 
Furthermore, the comparison with the functional k-means settings introduced by 
\citet{Abraham2003} and \citet{Chiou2007} indicates our methodology 
as an attractive alternative.

In Section~\ref{sec:application} we used our approach to analyze a set 
of drosophila melanogaster gene expression data. Our results 
appear particularly encouraging since the adjusted rand index (ARI), computed
according to the classification of \citet{Arbeitman2002}, has been found 
approximately equal to 0.96. Moreover, our results are consistent with those 
obtained by \citet{Chiou2007} for the same dataset. 

Some aspects have not been investigated yet and will guide our 
further research. First, we believe that our proposal can be valuable also within 
clustering frameworks different from the k-means one. Second, in all the examples 
proposed here, we supposed a smooth signal describing the trend of the series. 
In some cases, for example by dealing with EEG or spectroscopy measurements, 
this hypothesis does not appear appropriate and different definitions of 
the P-spline penalty could be useful. 
Finally, P-splines are easily generalizable to analyze discrete observations and 
multidimensional (e.g. defined over space and time) measurements 
\citep[see e.g.][]{Marx2005, Eilers1996, Eilers2010}. We 
have intention to investigate the possibility to extend the 
P-spline based k-means procedure in order to deal with such data.
%
\section*{Acknowledgments}
Gianluca Frasso acknowledges financial support from IAP
research network P7/06 of the Belgian Government (Belgian Science Policy).

\end{document}